\documentclass[aps, twocolumn, prx, longbibliography]{revtex4-1}
\usepackage{amsfonts}
\usepackage{amsmath}
\usepackage{amssymb}
\usepackage{mathptmx} 
\usepackage{graphicx}
\usepackage{bm}
\usepackage{natbib}
\usepackage{color}
\usepackage[colorlinks=true,linkcolor=blue,anchorcolor=red,citecolor=blue, urlcolor=blue]{hyperref}
\usepackage{ulem}

\begin{document}
\title{Impact of epitaxial strain on the topological-nontopological phase diagram and semimetallic behavior of InAs/GaSb composite quantum wells}
\author{H.~Irie}\email{hiroshi.irie.ke@hco.ntt.co.jp}\thanks{These authors contributed equally to this work.}
\author{T.~Akiho}\thanks{These authors contributed equally to this work.}
\author{F.~Cou\"{e}do}\altaffiliation[Present address: ]{Laboratoire National de M\'{e}trologie et d'Essais (LNE) Quantum Electrical Metrology Department, Avenue Roger Hennequin, 78197 Trappes, France}
\author{R.~Ohana}
\author{K.~Suzuki}\altaffiliation[Present address: ]{Fukuoka Institute of Technology, Fukuoka 811-0295, Japan}
\author{K.~Onomitsu}
\author{K.~Muraki}
\affiliation{NTT Basic Research Laboratories, NTT Corporation, 3-1 Morinosato-Wakamiya, Atsugi 243-0198, Japan}
\pacs{PACS number}
\date{\today}

\begin{abstract}
We study the influence of epitaxial strain on the electronic properties of InAs/GaSb composite quantum wells (CQWs), host structures for quantum spin Hall insulators, by transport measurements and eight-band $\mathbf{k\cdot p}$ calculations.
Using different substrates and buffer layer structures for crystal growth, we prepare two types of samples with vastly different strain conditions.
CQWs with a nearly strain-free GaSb layer exhibit a resistance peak at the charge neutrality point that reflects the opening of a topological gap in the band-inverted regime.
In contrast, for CQWs with 0.50\% biaxial tensile strain in the GaSb layer, semimetallic behavior indicating a gap closure is found for the same degree of band inversion.
Additionally, with the tensile strain, the boundary between the topological and nontopological regimes is located at a larger InAs thickness.
Eight-band $\mathbf{k\cdot p}$ calculations reveal that tensile strain in GaSb not only shifts the phase boundary but also significantly modifies the band structure, which can result in the closure of an indirect gap and make the system semimetallic even in the topological regime.
Our results thus provide a global picture of the topological-nontopological phase diagram as a function of layer thicknesses and strain.
\end{abstract}
\maketitle

\section{Introduction}

InAs/GaSb composite quantum wells (CQWs), in which electrons and holes are separately confined in the InAs and GaSb wells, have drawn renewed attention as a host structure for quantum spin Hall insulators (QSHIs), or two-dimensional topological insulators~\cite{Liu2008, Knez2010, Knez2011, Suzuki2013, Du2015, Suzuki2015, Qu2015, Couedo2016, Karalic2016, Nichele2014}.
A CQW, usually flanked by Al$_{x}$Ga$_{1-x}$Sb barriers ($x = 0.5$--$1.0$), is characterized by a broken-gap type-II band alignment where the conduction-band bottom of InAs is located below the valence-band top of GaSb.
When the thicknesses of the InAs and GaSb layers are such that the quantum confinement is not too strong, the CQW has an inverted band structure; namely, the first electron subband lies below the first heavy-hole subband at the Brillouin zone center.
In the absence of coupling between electrons and holes, the system becomes a semimetal, with the in-plane dispersion curves of the electron and hole subbands intersecting each other at a finite momentum.
However, in the presence of finite coupling between the electron and hole wave functions, a hybridization gap of a few meV opens at the band crossing point.
This energy gap, possessing a nature distinct from that of normal semiconductors, gives rise to topologically protected gapless states at the sample edges~\cite{Liu2008}.

Although the existence of the hybridization gap in InAs/GaSb CQWs with the inverted band order was confirmed in early experiments~\cite{Yang1997,Lakrimi1997}, subsequent transport measurements have consistently shown that the conductivity remains finite at low temperatures even when the Fermi level is adjusted to the middle of the gap~\cite{Knez2010,Cooper1998}.
Since this residual bulk conductivity is in most cases comparable to or greater than the conductance $e^{2}/h$ expected for the edge states ($e$ is the elementary charge and $h$ is Planck's constant), it represents an important issue in exploring and exploiting the exotic properties of the topological edge states.
While disorder-induced in-gap states are a likely cause, experiments show that the residual conduction remains strong even in high-quality samples~\cite{Qu2015, Karalic2016} and tends to weaken in more disordered samples~\cite{Du2015, Charpentier2013}.
It has been argued on the basis of quantum transport theory that the low-temperature conductivity in the hybridization gap of an electron-hole coupled system has an intrinsic lower limit determined by the degree of band inversion and the magnitude of the hybridization gap~\cite{Naveh2001}.
However, as noted in many studies, the semimetallic behavior of the conductivity, represented by the absence of activated temperature dependence, suggests an effective closure of the band gap, with its origin being discussed in terms of the anisotropy, or warping, of the valence band~\cite{Lakrimi1997, Suzuki2015, De-Leon1999}.

Another factor of particular notice is the epitaxial strain that arises from the lattice mismatch among constituent materials of heterostructures, which has recently attracted interest as a useful tool to engineer the band structure of QSHIs.
Specifically, a lattice-mismatched system of InAs/In$_{x}$Ga$_{1-x}$Sb CQWs, in which the In$_{x}$Ga$_{1-x}$Sb layer is under compressive epitaxial strain, has been shown to have an enlarged hybridization gap and thus exhibits significantly reduced residual conduction~\cite{Akiho2016, Du2017, Li2017}.
These results remind us of the need to fully take into account the strain effects also in the conventional InAs/GaSb/AlSb system~\cite{Zakharova2002, Jiang2017}, termed ``the 6.1-\AA~family"~\cite{Kroemer2004}, that is generally thought of as being approximately lattice-matched.
Indeed, Zakharova \textit{et al.} have calculated the band structures of InAs/GaSb CQWs pseudomorphically grown on InAs and GaSb substrates and compared them with that for the unstrained case~\cite{Zakharova2002}.
They showed that strain strongly affects the order of the levels at the Brillouin zone center, subband dispersion, and magnitude of the hybridization gap.
By noting that the band anisotropy depends on strain, they also showed that, when the InAs well is wide, the band gap becomes negative and indirect (i.e., semimetallic) for a structure grown on a GaSb substrate, whereas it is positive and direct for the same structure grown on an InAs substrate.
On the experimental side, externally applied hydrostatic pressure~\cite{Beerens1987} and uniaxial strain~\cite{Tiemann2017} have been shown to induce measurable changes in the electronic properties of the InAs/GaSb system.

In this paper, we study the influence of epitaxial strain on the electronic properties of InAs/GaSb CQWs via magnetotransport measurements and theoretical calculations.
In particular, we focus on the effects of biaxial tensile strain in the GaSb layer, which become relevant in some situations but have not been considered in previous studies.
By utilizing different substrates and buffer layer structures for crystal growth, a tensile strain of up to $\sim 0.50$\% ($\sim 1.13$\%) is exerted on the GaSb (InAs) layer via the lattice-constant mismatch between the quantum-well (QW) layers and the underlying buffer layer.
Analyzing the magnetotransport data based on an electrostatic capacitor model of the CQWs enables us to distinguish the inverted and non-inverted band alignment.
A comparison of samples with varying QW thickness and different strain conditions reveals that, for the same GaSb thickness, the tensile strain shifts the boundary between the inverted and non-inverted regimes to a larger InAs thickness.
For a thick InAs well, a CQW with tensile-strained GaSb shows semimetallic behavior indicating a gap closure, whereas an unstrained-GaSb CQW with a similar degree of band inversion shows a gap.
Eight-band $\mathbf{k\cdot p}$ calculations also reveal that the tensile strain makes the phase diagram as a function of InAs and GaSb layer thicknesses essentially different from that known for the unstrained case.
Our results will be useful to better understand previous reports on InAs/GaSb CQWs, where the strain effects have often been overlooked, and appropriately design future experiments.

\section{Experiment}

\subsection{Heterostructure design}

\begin{figure}[ptb]
\includegraphics{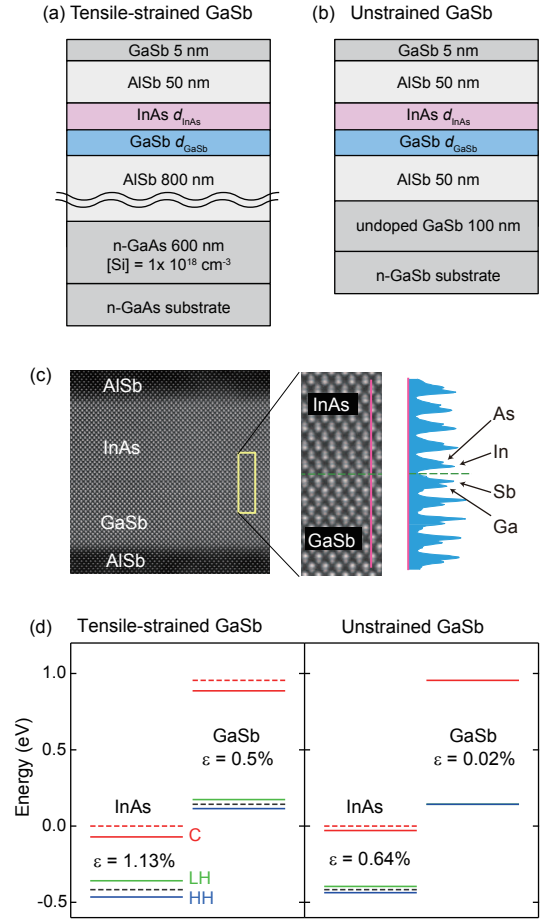}
\caption{\label{Fig1}(color online) (a), (b) Heterostructures using two types of buffer layer sequence referred to as (a) tensile-strained- and (b) unstrained-GaSb CQWs in the main text.
(c) HAADF-STEM of a tensile-strained-GaSb CQW.
The intensity plot (right panel) taken along the magenta bar on the enlarged image (middle panel) is used to determine the layer number.
(d) Conduction and valence band edges for (left) tensile-strained- and (right) unstrained-GaSb CQWs.
Dashed lines represent band edges of InAs and GaSb under no strain.}
\end{figure}

We study two types of CQWs with different buffer layer structures grown by molecular beam epitaxy.
The first type, hereafter denoted as the ``tensile-strained-GaSb" CQW, was formed on a thick (800 nm) AlSb buffer layer grown on a GaAs substrate [Fig.~\ref{Fig1}(a)].
Because of the large lattice mismatch ($\sim 8$\%) between GaAs ($a_\mathrm{GaAs} = 5.65325$ \AA) and AlSb ($a_\mathrm{AlSb} = 6.1355$ \AA), a nearly complete strain relaxation occurs in the AlSb layer near the AlSb/GaAs interface.
Since the lattice constants of both InAs ($a_\mathrm{InAs} = 6.0583$ \AA) and GaSb ($a_\mathrm{GaSb} = 6.0959$ \AA) are smaller than $a_\mathrm{AlSb}$, both the InAs and GaSb layers comprising the CQW are tensile strained.
For the growth on GaAs substrates, the use of a thick (800 nm) AlSb buffer layer was essential to reduce the threading dislocation density and obtain CQWs of reasonable quality~\cite{Cooper1998}.
The second type, denoted as the ``unstrained-GaSb" CQW, was formed on a thin (50 nm) AlSb buffer layer grown on a GaSb substrate [Fig.~\ref{Fig1}(b)].
Because of the small thickness of the AlSb buffer layer and the rather small lattice mismatch ($\sim 0.6$\%) between GaSb and AlSb, the AlSb buffer layer remains fully strained, with its in-plane lattice constant equal to $a_\mathrm{GaSb}$.
Accordingly, the GaSb layer of the CQW formed on it is unstrained.
The InAs layer, on the other hand, is tensile-strained, due to the $\sim 0.6$\% lattice mismatch between InAs and GaSb.

\begin{table*}
\caption {In-plane lattice constant of AlSb buffer layer ($a_\mathrm{buffer}$) measured by HRXRD and resultant strain in GaSb and InAs QW layers ($\varepsilon_\mathrm{GaSb}$ and $\varepsilon_\mathrm{InAs}$).
$a_\mathrm{buffer}$ is obtained as the average of RSM measurements using orthogonal incident angles ($[110]$ and $[1 \overline{1} 0]$).
The relaxation ratio is defined as $(a_\mathrm{buffer} - a_\mathrm{sub})/(a_\mathrm{AlSb} - a_\mathrm{sub})$, where $a_\mathrm{sub}$ is the lattice constant of the substrate, i.e., $a_\mathrm{GaAs}$ for tensile-strained-GaSb samples and $a_\mathrm{GaSb}$ for unstrained-GaSb samples, respectively.}
\begin{ruledtabular}
\begin{tabular}{ccccc}
Sample  &  $a_\mathrm{buffer}$ (\AA)  &  Relaxation ratio (\%)  &  $\varepsilon_\mathrm{GaSb}$ (\%)  &  $\varepsilon_\mathrm{InAs}$ (\%)\\
\hline
Tensile-strained GaSb & $6.1266$ & $98.15$ & $0.50$ & $1.13$ \\
Unstrained GaSb & $6.0972$ & $3.3$ & $0.02$ & $0.64$ \\
\end{tabular}
\end{ruledtabular}
\end{table*}

In both types of CQWs, the thickness $d_\mathrm{GaSb}$ of the GaSb layer comprising the CQW was fixed at $7.3$~nm, which corresponds to 24 monolayers (MLs).
The thickness $d_\mathrm{InAs}$ of the InAs was varied from $9.1$ to $11.8$~nm (30 to 39 MLs) for the tensile-strained-GaSb CQWs and from $9.1$ to $10.6$~nm (30 to 35~MLs) for the unstrained-GaSb CQWs.
We calibrated the layer thickness by counting the number of atomic layers in high-angle annular dark field scanning transmission electron microscopy (HAADF-STEM) images with atomic resolution [Fig.~\ref{Fig1}(c)].

The magnitudes of strain in the CQWs were evaluated using two-dimensional reciprocal space mapping (RSM) of the high-resolution x-ray diffraction (HRXRD) (see Appendix A for details).
The in-plane lattice constant, relaxation ratio, and strain calculated from the mean value of the measurements for $[110]$ and $[1 \overline{1} 0]$ directions are summarized in Table~I.
In the tensile-strained-GaSb samples, the InAs and GaSb layers in the CQWs are 1.13 and 0.50\% tensile-strained, respectively.
For the unstrained-GaSb CQWs, the InAs layer is 0.64\% tensile-strained, while the GaSb layer is nearly unstrained.

Figure \ref{Fig1}(d) shows the band-edge alignment of InAs and GaSb for the cases of tensile-strained- and unstrained-GaSb CQWs.
The dashed lines represent the positions of the band edges for the case where no strain is taken into account.
With no strain, the conduction-band bottom of InAs is 0.14 eV lower than the valence-band top of GaSb.
With tensile strain in the InAs layer, its conduction-band bottom shifts to lower energy.
With tensile strain in the GaSb layer, its valence-band top is split into heavy-hole (HH) and light-hole (LH) bands.
It is worth noting that the tensile strain lowers the HH band with respect to the LH band.
The energy difference between the bulk band edges of the HH and conduction bands barely depends on the strain.
However, as we will elaborate in section III, the band overlap between the electron and HH subbands in a CQW depends on the strain, due to the mixing between the electron and LH subbands.

\subsection{Magnetotransport and equivalent-circuit analysis}

Here, we describe the procedure to characterize the electronic properties of CQWs.
We employed Hall-bar devices with a length and width of 180 and 50~$\mu$m, respectively.
For this large device size, the contribution of edge conduction is negligible for most of the cases studied here, so the measured resistance reflects the bulk property.
We measured the electronic properties at 2 K under a perpendicular magnetic field $B$ up to 14 T using the standard lock-in technique.
The Hall-bar devices are fitted with a front gate, which we use to tune the Fermi level across the charge neutrality point (CNP).
The gate insulator was 40-nm-thick atomic-layer-deposited aluminum oxide.
The front-gate voltage $V_\mathrm{FG}$ was swept in the range $-1.5$~V $\leq V_\mathrm{FG} \leq$ 3.5 V. The lowest $V_\mathrm{FG}$ was limited to $-1.5$~V, because hysteresis occurs at $V_\mathrm{FG} <$ $-1.5$~V, which shifts the device characteristics.
All the data presented in this paper were taken with the substrate (back gate) kept at 0 V.
In this subsection, we outline our analysis using data taken from a tensile-strained-GaSb CQW with $d_\mathrm{InAs} = 11.8$ nm.

\begin{figure*}[ptb]
\includegraphics{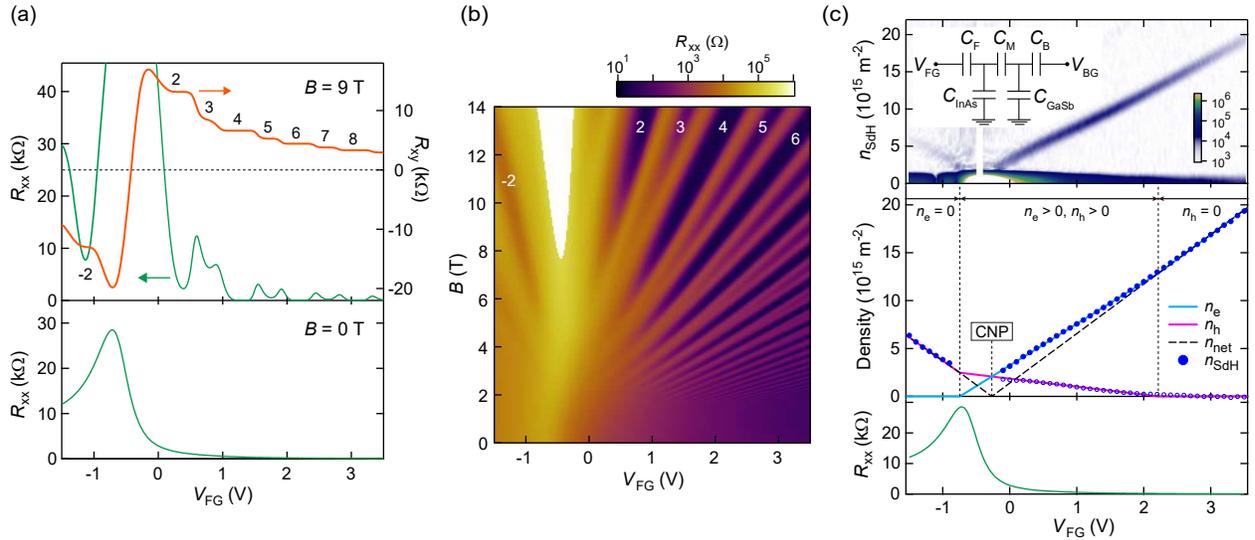}
\caption{(color online) Transport properties of tensile-strained-GaSb CQW with $d_\mathrm{InAs} = 11.8$ nm at $T = 2$ K.
(a) Longitudinal resistance $R_{xx}$ at $B = 0$ and 9 T, and Hall resistance $R_{xy}$ at $B = 9$ T.
Landau-level filling factor derived from the $R_{xy}$ plateau values is shown.
(b) Color plot of $R_{xx}$ as a function of $V_\mathrm{FG}$ and $B$.
Filling factor derived from $R_{xy}$ is shown.
(c) Top: Fast Fourier transform (FFT) spectra of $R_{xx}(1/B)$ performed at each $V_\mathrm{FG}$.
The vertical axis is the oscillation frequency multiplied by $2e/h$ to associate the peak position with the carrier density.
The inset shows the equivalent circuit used for analyzing the density change as a function of $V_\mathrm{FG}$.
Middle: $V_\mathrm{FG}$ dependence of carrier density.
The solid symbols represent the density obtained from the FFT spectra ($n_\mathrm{SdH}$).
The open symbols represent the difference between $n_\mathrm{SdH}$ and the linear fitting of $n_\mathrm{SdH}$ in the deep electron regime.
The lines are the electron density $n_\mathrm{e}$ (cyan), the hole density $n_\mathrm{h}$ (magenta), and the net carrier density $n_\mathrm{net}$ (black dashed), all calculated from the equivalent-circuit model.
Bottom: $R_{xx}$ vs $V_\mathrm{FG}$ at $B = 0$ T [same data as in (a)] replotted for comparison with the circuit-model analysis of the carrier density change with $V_\mathrm{FG}$.}
\label{Fig2}
\end{figure*}

Figure \ref{Fig2}(a) shows the dependence of the longitudinal resistance $R_{xx}$ at $B = 0$ and 9 T and the Hall resistance $R_{xy}$ at $B = 9$ T on the front-gate voltage $V_\mathrm{FG}$.
At $B = 0$ T, $R_{xx}$ exhibits a single peak at $V_\mathrm{FG} = -0.72$~V without any additional features.
As we will show below, this peak is not located at the CNP and therefore not a manifestation of an energy gap opening.
$R_{xy}$ changes sign at $V_\mathrm{FG} = -0.4$~V, demonstrating that the majority carrier type changes from holes to electrons with increasing $V_\mathrm{FG}$.
The fact that the sign change occurs through $R_{xy} = 0$ (instead of diverging to $\pm \infty$) indicates that electrons and holes coexist over a range of $V_\mathrm{FG}$.
At $B = 9$ T, several $R_{xy}$ plateaus along with $R_{xx}$ dips due to quantum Hall effects are observed in both the electron- and hole-dominant regions.

Figure \ref{Fig2}(b) shows a color plot of $R_{xx}$ as a function of $V_\mathrm{FG}$ and $B$.
Shubnikov-de Haas (SdH) oscillations in the low-$B$ regime evolve into quantum Hall effects with vanishing $R_{xx}$ in the high-$B$ regime.
We deduced the carrier densities as a function of $V_\mathrm{FG}$ by fast Fourier transform (FFT) analysis of the SdH oscillations at each $V_\mathrm{FG}$ with respect to $1/B$.
The FFT power spectra are shown in the top panel of Fig.~\ref{Fig2}(c) as a color plot.
The vertical axis is the carrier density $n_\mathrm{SdH}$, which is related to the frequency $f_{1/B}$ as $n_\mathrm{SdH} = g_\mathrm{s}(e/h)f_{1/B}$, where $g_\mathrm{s}$ is the spin degeneracy.
Throughout this paper, we take $g_\mathrm{s} = 2$, which gives results consistent with those obtained from $R_{xy}$.
The FFT analysis identifies only one frequency (i.e., one density $n_\mathrm{SdH}$) at each $V_\mathrm{FG}$, which is plotted as solid circles in the middle panel of Fig.~\ref{Fig2}(c).
$n_\mathrm{SdH}$ first decreases with $V_\mathrm{FG}$ at $V_\mathrm{FG} \le -0.9$~V and then starts to increase at $V_\mathrm{FG} \ge -0.1$~V.
This $V_\mathrm{FG}$ dependence confirms that $n_\mathrm{SdH}$ represents the density of majority carriers at each $V_\mathrm{FG}$, i.e., hole density $n_\mathrm{h}$ at $V_\mathrm{FG} \le -0.9$~V and electron density $n_\mathrm{e}$ at $V_\mathrm{FG} \ge -0.1$~V.

We determine both $n_\mathrm{e}$ and $n_\mathrm{h}$ at each $V_\mathrm{FG}$ by analyzing the $V_\mathrm{FG}$ dependence of $n_\mathrm{SdH}$ using the equivalent-circuit model illustrated in the inset of Fig.~\ref{Fig2}(c) (top panel)~\cite{Qu2015, Akiho2016}.
In addition to the geometrical capacitances that represent the couplings to the front and back gates ($C_\mathrm{F}$ and $C_\mathrm{B}$) and between the InAs and GaSb QW layers ($C_\mathrm{M}$), the model includes the quantum capacitances of the QW layers--- $C_\mathrm{InAs} = e^{2} m^{*}_\mathrm{e,InAs}/\pi \hbar^{2}$ and $C_\mathrm{GaSb} = e^{2} m^{*}_\mathrm{h,GaSb}/\pi \hbar^{2}$ with $m^{*}_\mathrm{e, InAs}$ ($m^{*}_\mathrm{h,GaSb}$) the effective mass of electrons in InAs (holes in GaSb) and $\hbar = h/2\pi$.
For simplicity, we neglect the energy dependence of the effective masses and the hybridization between the electron and hole bands.
With this circuit model, $n_\mathrm{e}$ and $n_\mathrm{h}$ are obtained as the charges stored in $C_\mathrm{InAs}$ and $C_\mathrm{GaSb}$, respectively.
We chose the $m^{*}_\mathrm{e,InAs}$ and $m^{*}_\mathrm{h,GaSb}$ values that give the best fit to the observed $V_\mathrm{FG}$ dependence of $n_\mathrm{SdH}$.
(See Appendix B for the details of the fitting procedure and the parameters used in the analysis.)

In the middle panel of Fig.~\ref{Fig2}(c), we plot the calculated $n_\mathrm{e}$ and $n_\mathrm{h}$ (solid lines) along with the net carrier density defined as $n_\mathrm{net} = |n_\mathrm{e} - n_\mathrm{h}|$ (dashed line).
The model reproduces the $V_\mathrm{FG}$ dependence of $n_\mathrm{SdH}$ over the entire range, including the slope change at $V_\mathrm{FG} = 2.2$~V.
As shown by the magenta solid line, $n_\mathrm{h}$ decreases to $0$ at $V_\mathrm{FG} = 2.2$~V, from which the slope change in $n_\mathrm{SdH}$ is understood as arising from the onset of the hole-band occupation.
This, in turn, allows one to locate the boundary between the single-carrier regime ($n_\mathrm{e} > 0$ and $n_\mathrm{h} = 0$) and the electron-hole coexistence regime ($n_\mathrm{e}, n_\mathrm{h} > 0$).
For $n_\mathrm{e} > n_\mathrm{h}$, the hole density can be expressed as $n_\mathrm{h} = n_\mathrm{e} - n_\mathrm{net}$.
We experimentally deduced the hole density in the coexistence regime by substituting $n_\mathrm{e}$ and $n_\mathrm{net}$ in this equation with the measured $n_\mathrm{SdH}$ and its linear fit in the single-carrier regime extrapolated to the coexistence regime, respectively.
The latter is justified because the slope of $n_\mathrm{net}$ remains unchanged in the single-carrier and coexistence regimes.
The hole density deduced in this way, shown as open symbols in the middle panel of Fig.~\ref{Fig2}(c), agrees with the calculated $n_\mathrm{h}$ (magenta solid line).
Similarly, by extrapolating $n_\mathrm{SdH}$ in the coexistence regime to $0$, the onset of the electron-band occupation is found to be at $V_\mathrm{FG} = -0.74$~V.
The coexistence of electrons and holes over a finite $V_\mathrm{FG}$ range (from $-0.74$ to $2.2$~V) demonstrates that this CQW has an inverted band structure.
One can locate the CNP at $V_\mathrm{FG} = -0.27$~V where $n_\mathrm{e}$ and $n_\mathrm{h}$ cross.
The density $n_\mathrm{cross}$ (= $2.1 \times 10^{15}$ m$^{-2}$) at this crossing point provides a quantitative measure of the degree of band inversion.
$n_\mathrm{net}$ ($= |n_\mathrm{e} - n_\mathrm{h}|$) shows a V-shaped $V_\mathrm{FG}$ dependence, with a constant slope of $5.2 \times 10^{15}$ m$^{-2}$V$^{-1}$, which reflects mostly the geometrical capacitance to the front gate.

Turning to the $R_{xx}$ vs $V_\mathrm{FG}$ curve at $B = 0$ T, which we replot in the bottom panel of Fig.~\ref{Fig2}(c) for comparison, we notice that the $R_{xx}$ peak is not located at the CNP ($V_\mathrm{FG} = -0.27$~V).
Rather, the peak position is close to the onset of the electron-band occupation ($V_\mathrm{FG} = -0.74$~V).
This is reasonable as the electrons have higher mobility than holes.
The classical two-carrier-model analysis of magneto-conductance using the $n_\mathrm{e}$ and $n_\mathrm{h}$ values obtained above (not shown) gives a mobility ratio of 4 near the CNP.
Importantly, $R_{xx}$ shows no feature at the CNP.
As we discuss in section III, this is due to the semimetallic band structure caused by the tensile strain in the GaSb layer.

\subsection{Effects of InAs thickness and buffer layer structure}

\subsubsection*{$d_\mathrm{InAs}$ dependence of tensile-strained-GaSb CQWs}

\begin{figure*}[ptb]
\includegraphics{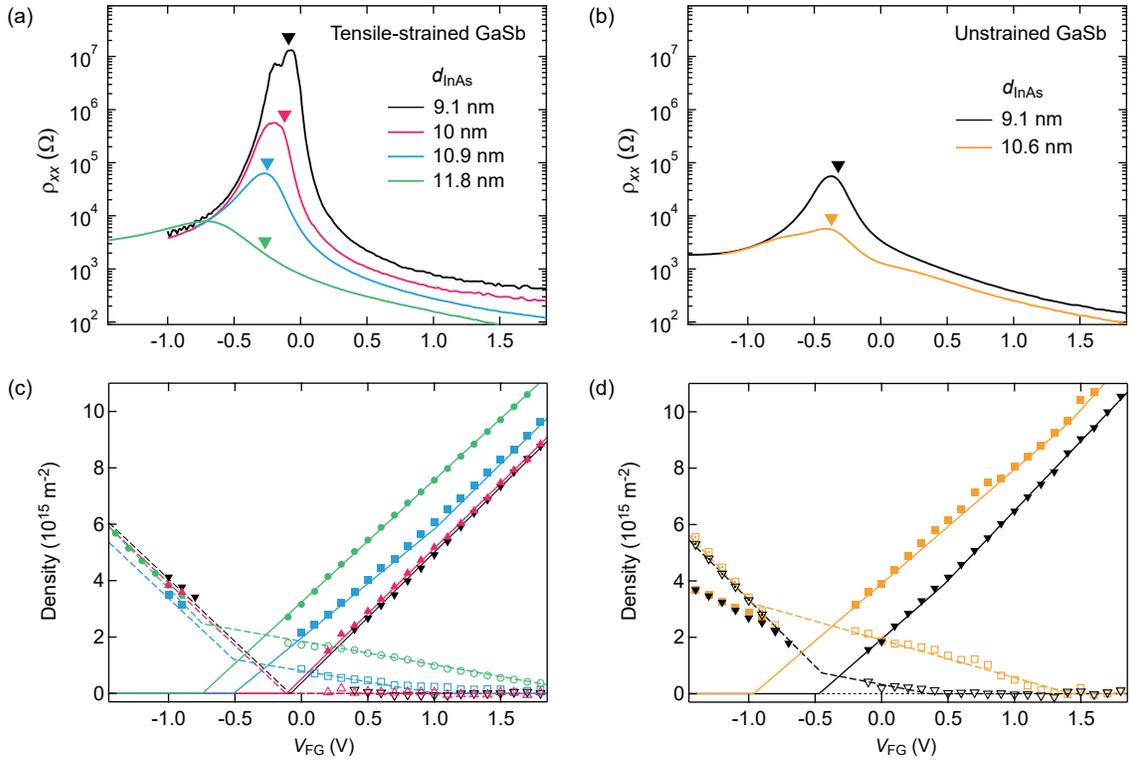}
\caption{(color online) Impacts of the InAs layer thickness and buffer layer structures on transport properties of InAs/GaSb CQWs.
(a), (b) Sheet resistivity as a function of $V_\mathrm{FG}$ for (a) tensile-strained-GaSb and (b) unstrained-GaSb CQWs measured at $T = 2$~K and $B = 0$~T.
The arrows represent the position of the charge neutrality point determined from the analysis shown in (c) and (d).
(c), (d) Carrier density obtained from the FFT analysis of the SdH oscillations.
Solid symbols represent the peak position ($n_\mathrm{SdH}$) of the FFT spectra.
Open symbols are the difference between $n_\mathrm{SdH}$ and the linear fit of $n_\mathrm{SdH}$ extrapolated from the deep electron regime.
Open symbols with a center dot in (d) are obtained from the magnetic field position of the $\nu = -2$ quantum Hall state.
The solid (dashed) lines are electron (hole) density calculated from the equivalent-circuit model.}
\label{Fig3}
\end{figure*}

Using the procedure outlined in the previous subsection, we first examine the $d_\mathrm{InAs}$ dependence of tensile-strained-GaSb CQWs.
Figure \ref{Fig3}(a) shows the $B = 0$ sheet resistivity $\rho_{xx}$ of the tensile-strained-GaSb CQWs with different $d_\mathrm{InAs}$.
In all samples, $\rho_{xx}$ exhibits a single peak, with the height monotonically increasing with decreasing $d_\mathrm{InAs}$.
We performed magnetotransport measurements and analyses for CQWs with $d_\mathrm{InAs} = 9.1$, $10$, and $10.9$~nm, similarly to what we did for the CQW with $d_\mathrm{InAs} = 11.8$ nm shown in Fig.~\ref{Fig2}.
Figure \ref{Fig3}(c) compiles the results for the four CQWs, where we plot $n_\mathrm{SdH}$ vs $V_\mathrm{FG}$ obtained from the FFT analysis (solid symbols) along with $n_\mathrm{e}$ (solid lines) and $n_\mathrm{h}$ (dashed lines) calculated using the equivalent-circuit model (parameters are summarized in Appendix B).
The hole density deduced from the SdH data following the procedure described in the previous subsection is shown as open symbols.
While the analysis used the SdH data taken over a wide $V_\mathrm{FG}$ range up to 3.5~V as in Fig.~\ref{Fig2}(c), we only show the results for $V_\mathrm{FG} \le 1.85$~V in Fig.~\ref{Fig3}(c) to highlight the behavior near the CNP.

Like the CQW with $d_\mathrm{InAs} = 11.8$ nm, $n_\mathrm{SdH}$ vs $V_\mathrm{FG}$ of the one with $d_\mathrm{InAs} = 10.9$ nm shows a slope change at $V_\mathrm{FG} = 0.9$~V due to hole-band occupation, indicating an inverted band structure.
The CQWs with $d_\mathrm{InAs} = 10.9$ and $11.8$ nm have $n_\mathrm{cross}$ of $1.0$ and $2.1 \times 10^{15}$ m$^{-2}$, respectively.
The smaller $n_\mathrm{cross}$ for $d_\mathrm{InAs} = 10.9$ nm indicates a shallower band inversion that results from the stronger quantum confinement of electrons in InAs.
In contrast, the $n_\mathrm{SdH}$ vs $V_\mathrm{FG}$ for $d_\mathrm{InAs} = 9.1$ and $10$ nm exhibits a simple V shape, with no slope change, indicating no electron-hole coexistence.
This indicates that these CQWs are in the non-inverted regime, as a result of the even stronger quantum confinement in the InAs QW.

The equivalent-circuit analysis also provides the position of the CNP, which is marked by arrows in Fig.~\ref{Fig3}(a).
In the CQWs with $d_\mathrm{InAs} = 9.1$ and $10$~nm, the $\rho_{xx}$ peak position roughly coincides with the CNP, as expected for the normal semiconducting gap.
For $d_\mathrm{InAs} = 10$~nm, the $\rho_{xx}$ peak is slightly shifted to the hole regime, presumably due to the large mobility difference between electrons and holes.
For the CQW with $d_\mathrm{InAs} = 10.9$~nm, despite its inverted band structure, the $\rho_{xx}$ peak position coincides with the CNP.
This contrast with the one with $d_\mathrm{InAs} = 11.8$ nm, indicating a gap opening.

\subsubsection*{$d_\mathrm{InAs}$ dependence of unstrained-GaSb CQWs}

\begin{figure}[ptb]
\includegraphics{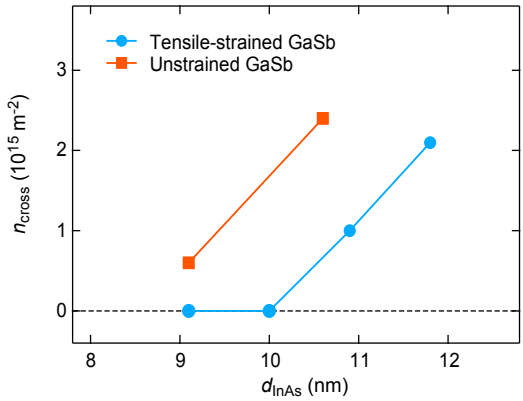}
\caption{(color online) Dependence of $n_\mathrm{cross}$ on $d_\mathrm{InAs}$ for the tensile-strained- and unstrained-GaSb CQWs, determined from the data in Figs.~\ref{Fig3}(c) and \ref{Fig3}(d).
Data points with $n_\mathrm{cross} = 0$ represent CQWs in the non-inverted regime.}
\label{Fig4}
\end{figure}

Next, we examine the properties of the unstrained-GaSb CQWs, which are shown in Figs.~\ref{Fig3}(b) and \ref{Fig3}(d).
The carrier density plot in Fig.~\ref{Fig3}(d) demonstrates that the two studied samples ($d_\mathrm{InAs} = 9.1$ and $10.6$~nm) are both in the band-inverted regime and thus have nonzero $n_\mathrm{cross}$
\footnote{
As seen in Fig.~\ref{Fig3}(d), the FFT analysis of the SdH oscillations in the hole-dominant regime yields unexpectedly low hole density (solid symbols) which cannot be explained by the equivalent-circuit model.
To circumvent this issue, we determined the hole density from the magnetic field position of the $\rho_{xx}$ minimum corresponding to the $\nu = -2$ quantum Hall state.
The hole density determined in this way [open symbols with a center dot in Fig.~\ref{Fig3}(d)] agrees with the behavior expected from the equivalent-circuit model.
The reason why the SdH oscillations yield a smaller hole density is unknown at present.
}.
Note that the tensile-strained-GaSb CQWs with similar InAs thickness ($d_\mathrm{InAs} = 9.1$ and $10.0$~nm) are non-inverted, which indicates that the band overlap is larger in the unstrained-GaSb CQWs.
This becomes clear by plotting $n_\mathrm{cross}$ as a function of $d_\mathrm{InAs}$ (Fig.~\ref{Fig4}).
The larger $n_\mathrm{cross}$ for the same $d_\mathrm{InAs}$ demonstrates a larger degree of band inversion in the unstrained-GaSb CQWs.
Accordingly, the boundary between inverted ($n_\mathrm{cross} > 0$) and non-inverted ($n_\mathrm{cross} = 0$) regimes is located at a larger $d_\mathrm{InAs}$ in the tensile-strained-GaSb CQWs.

Turning back to the transport data in Fig.~\ref{Fig3}(b), we find that in both samples the position of the $\rho_{xx}$ peak matches the CNP.
We note in particular that this holds even in the deeply inverted $d_\mathrm{InAs} = 10.6$~nm CQW with large $n_\mathrm{cross}$ of $2.4 \times 10^{15}$~m$^{-2}$.
This shows that the observed $\rho_{xx}$ peak originates from an energy gap, thus corroborating the opening of a hybridization gap.
This contrasts with the case of tensile-strained-GaSb CQW with $d_\mathrm{InAs} = 11.8$~nm, where a semimetallic behavior is observed for the similar $n_\mathrm{cross}$ of $2.1 \times 10^{15}$~m$^{-2}$.
Furthermore, the CQW with $d_\mathrm{InAs} = 10.6$~nm has a satellite dip (or a shoulder) on both sides of the $\rho_{xx}$ peak [Fig.~\ref{Fig3}(b)].
Similar features have previously been reported for high-quality CQWs in the inverted regime~\cite{Knez2010, Qu2015, Karalic2016, Nguyen2015, Shojaei2018} and have been interpreted to be due to the van-Hove singularity in the density of states at the hybridization gap edges~\cite{Knez2010,De-Leon1999}.
The opening and absence of the hybridization gap in unstrained- and tensile-strained-GaSb CQWs with a similar degree of band inversion, revealed by the transport data in Fig.~\ref{Fig3}, are shown to be consistent with the band calculation in the next section.

\section{$\mathbf{k\cdot p}$ Calculation}
\subsection{Impact of tensile strain on the band structure}

To understand the difference in the electronic properties between tensile-strained- and unstrained-GaSb CQWs, we performed theoretical calculations based on the eight-band $\mathbf{k\cdot p}$ Hamiltonian with strain effects taken into account~\cite{Bir1974,Pollak1968} (material parameters are from Ref.~\cite{Vurgaftman2001}).
For simplicity, we use the axial approximation~\cite{Li2009}, neglecting the band anisotropy.
Furthermore, we assume a flat potential, neglecting the Hartree potential due to interlayer charge transfer.
Since the latter is known to be important for a quantitatively accurate description of the band structure in InAs/GaSb CQWs~\cite{Jiang2017,Andlauer2009}, our calculations should be taken as providing a qualitative guide.
The influence of the above approximations will be discussed later.

\begin{figure}[ptb]
\includegraphics{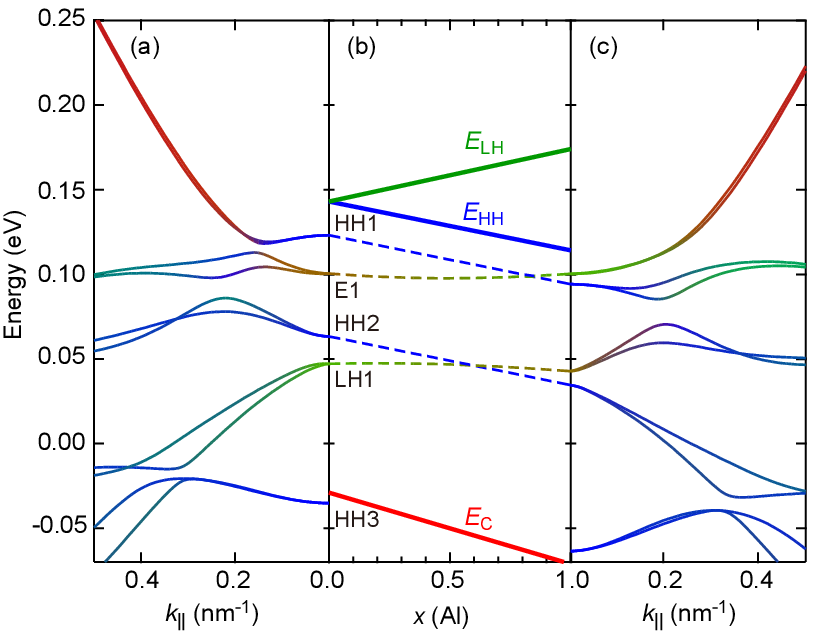}
\caption{(color online) (a), (c) Band dispersion of CQWs pseudomorphically grown on (a) GaSb and (c) AlSb substrates.
Layer thicknesses are $d_\mathrm{InAs} = 9$~nm and $d_\mathrm{GaSb} = 7.3$~nm.
The color of the lines denotes the band character; i.e., red, green, and blue represent an electron, light hole, and heavy hole character, respectively.
(b) Energy positions of bulk band edges and subbands as a function of the lattice constant represented by Al fraction $x$ of Al$_{x}$Ga$_{1-x}$Sb substrate.
Solid lines represent the band edge of conduction ($E_\mathrm{C}$), light-hole ($E_\mathrm{LH}$), and heavy hole ($E_\mathrm{HH}$) bands in the bulk.
Broken lines represent the subband levels at the zone center of E1, HH1, HH2, and LH1.
The energy is measured from the conduction band edge of InAs without strain.}
\label{Fig5}
\end{figure}

In Figs.~\ref{Fig5}(a) and \ref{Fig5}(c), we compare the band dispersions calculated for CQWs with the same layer thicknesses ($d_\mathrm{InAs} = 9$~nm and $d_\mathrm{GaSb} = 7.3$~nm) but grown pseudomorphically on (a) GaSb and (c) AlSb, which correspond to $\varepsilon_\mathrm{GaSb} = 0$ and 0.76\%, respectively.
The energies are plotted as a function of the magnitude $k_\parallel$ of the in-plane wave vector.
We use the line color to represent the character of the bands at each $k_\parallel$; namely, red, green, and blue indicate that the band predominantly has an electron (E), light hole (LH), and heavy hole (HH) character, respectively.

For the unstrained-GaSb case [Fig.~\ref{Fig5}(a)], the character of the bands at the Brillouin zone center ($k_\parallel = 0$) is identified as HH1, E1, HH2, LH1, and HH3 in descending order of energy.
(We hereafter label the bands according to their character at $k_\parallel = 0$.
Note that the actual band character varies with $k_\parallel$ and is therefore different from that at $k_\parallel = 0$.)
The inverted order of E1 and HH1, with the hybridization gap opening at around $k_\parallel = 0.15$~nm$^{-1}$ where E1 and HH1 bands anticross, shows that the system is in the topological regime.
Indeed, the band structure in Fig.~\ref{Fig5}(a) is typical of InAs/GaSb CQWs with an inverted band ordering~\cite{Liu2008}.
In contrast, the band structure for the tensile-strained-GaSb case, shown in Fig.~\ref{Fig5}(c), is obviously different.
The band located at the top has an E character at large $k_\parallel$, but acquires more of an LH character at small $k_\parallel$.
(We use the term ``band" to represent the pair of bands which are spin split at finite $k_\parallel$ but become degenerate at $k_\parallel = 0$.)
The HH1 band is located below it, which indicates that, despite the same layer thicknesses, the band order has changed by the tensile strain.
Note that the top of the HH1 band at $k_\parallel = 0.4$~nm$^{-1}$ (where the band has more of a LH character) is located above the bottom of the upper band.
Consequently, the system is a semimetal, even though these bands are separated by a gap at each $k_\parallel$~\footnote{The semimetallic band structure in Fig.~\ref{Fig5}(c) is similar to those in wide HgTe/CdTe QWs grown on a fully relaxed CdTe buffer layer, where the HgTe well is under 0.3\% biaxial tensile strain.
The coexistence of electrons and holes in such samples, revealed in transport measurements, has been interpreted to be due to the semimetallic band structure [Z. D. Kvon, E. B. Olshanetsky, D. A. Kozlov, N. N. Mikhailov, and S. A. Dvoretskii, JETP Letters \textbf{87,} 502 (2008);
E. Olshanetsky, Z. Kvon, N. Mikhailov, E. Novik, I. Parm, and S. Dvoretsky, Solid State Commun. \textbf{152,} 265 (2012);
P. Leubner, L. Lunczer, C. Br{\"{u}}ne, H. Buhmann, and L. W. Molenkamp, Phys. Rev. Lett. \textbf{117,} 086403 (2016).]}.

To see how the band crossover is caused by tensile strain, we plot the energy levels of the bands at $k_\parallel = 0$ in Fig.~\ref{Fig5}(b) (dashed lines) as a function of Al composition $x$ ($0 \le x \le 1$) of a virtual Al$_{x}$Ga$_{1-x}$Sb substrate that would produce $\varepsilon_\mathrm{GaSb}$ of 0 to 0.76\% ($\varepsilon_\mathrm{InAs}$ of 0.52 to 1.28\%).
The bulk band edges of InAs and GaSb are shown as thick solid lines.
With increasing $x$, the conduction-band edge of InAs ($E_\mathrm{C}$) shifts to lower energy as the tensile strain exerted on the InAs layer decreases its band gap via the deformation potential~\cite{Bir1974, Pollak1968}.
The valence-band edge of GaSb, which is fourfold degenerate at $x = 0$ (i.e., $\varepsilon_\mathrm{GaSb} = 0$), splits into LH and HH bands, with their energies, $E_\mathrm{LH}$ and $E_\mathrm{HH}$, shifting upward and downward, respectively.
While HH1 and HH2 (and HH3) levels follow the $x$ dependence of $E_\mathrm{HH}$, other levels, which have an E and LH character at $x = 0$, do not show a direct correspondence with either $E_\mathrm{C}$ or $E_\mathrm{LH}$.
We note that in heterostructures, HH bands are completely decoupled from other bands at $k_\parallel = 0,$ whereas finite coupling exists between E and LH bands even at $k_\parallel = 0$~\cite{Zakharova2001}.
In addition, a tensile strain shifts $E_\mathrm{C}$ downward and $E_\mathrm{LH}$ upward, which would bring the two levels identified as E1 and LH1 at $x = 0$ closer together if they are to follow $E_\mathrm{C}$ and $E_\mathrm{LH}$, respectively.
Consequently, E1 and LH1 levels are strongly mixed with increasing $x$ and, by $x = 1$, they almost swap their characters.
Because of this E1-LH1 mixing and the opposite $x$ dependence of $E_\mathrm{C}$ and $E_\mathrm{LH}$, the energies of the E1-LH1 mixed levels become barely dependent on $x$, leading to the level crossing with HH1 at $x = 0.8$ [Fig.~\ref{Fig5}(b)].
As seen in Fig.~\ref{Fig5}(c), at finite $k_\parallel$ the HH1 band mixes with the E1-LH1 band and loses the HH1 character with increasing $k_\parallel$.
Thus, the upturn of the HH1 band around $k_\parallel = 0.4$~nm$^{-1}$ (and the resultant semimetallic band structure) can be understood to arise from the HH-LH band character crossover and the tensile strain that lifts LH-like bands.

\subsection{InAs thickness dependence}

\begin{figure*}[ptb]
\includegraphics{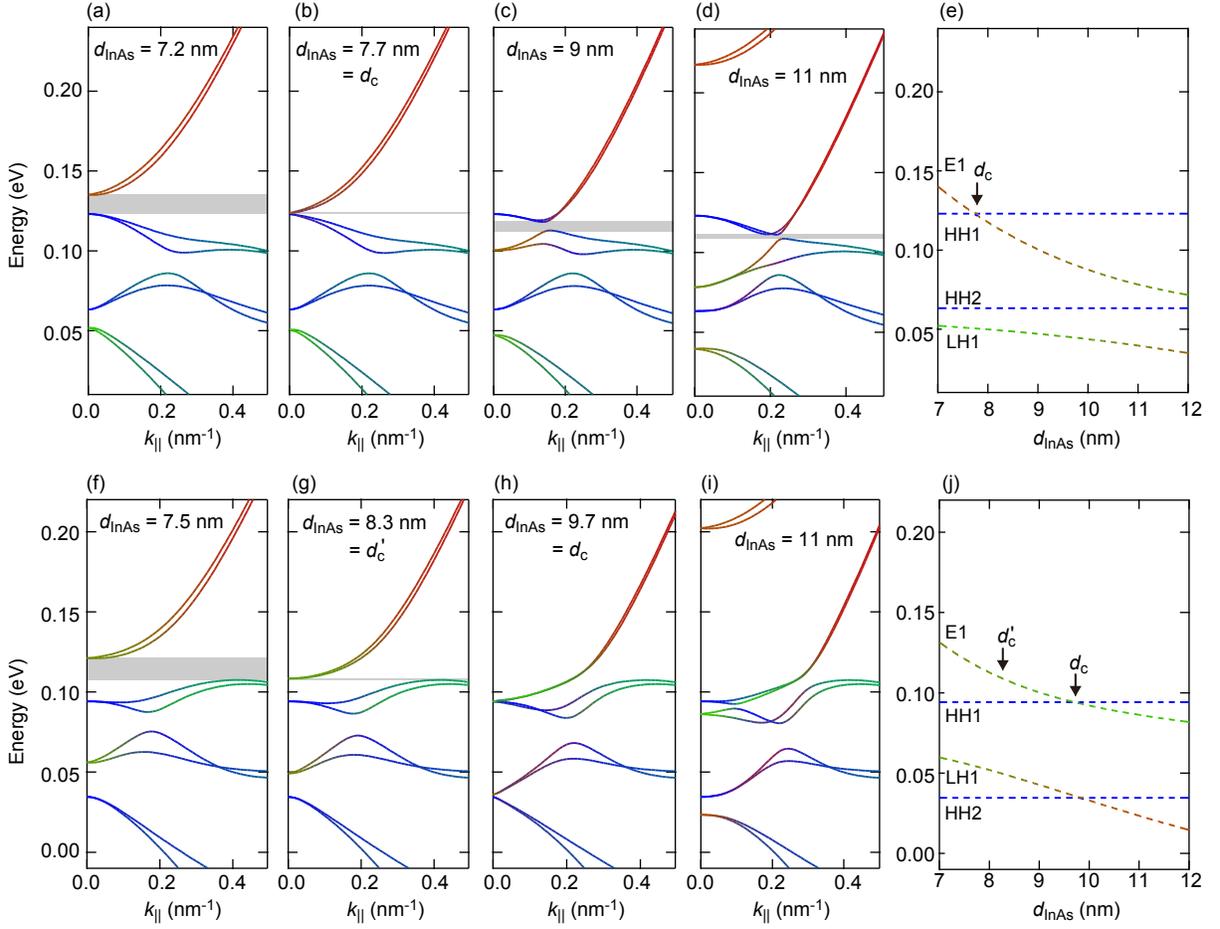}
\caption{(color online) (a)-(d) Band dispersion of CQWs pseudomorphically grown on a GaSb substrate for different $d_\mathrm{InAs}$ with $d_\mathrm{GaSb}$ fixed at 7.3~nm.
The grey shaded area represents the energy gap.
The color of the lines denotes the band character in the same way as that in Fig.~\ref{Fig5}.
(e) Energy levels at the Brillouin zone center of E1, HH1, HH2, and LH1 as a function of $d_\mathrm{InAs}$.
(f)-(i) Band dispersion of CQWs grown on AlSb substrate for different $d_\mathrm{InAs}$ with $d_\mathrm{GaSb}$ fixed at 7.3~nm.
(j) Subband levels at the zone center as a function of $d_\mathrm{InAs}$.}
\label{Fig6}
\end{figure*}

Next, we turn to the $d_\mathrm{InAs}$ dependence of the band dispersion.
Figure \ref{Fig6} shows how the band structure of CQWs grown on (a)-(d) GaSb and (f)-(i) AlSb substrates evolve when $d_\mathrm{InAs}$ is varied with $d_\mathrm{GaSb}$ fixed ($= 7.3$~nm).
As already shown in the previous subsection, the CQW grown on a GaSb substrate is in the topological regime at $d_\mathrm{InAs} = 9.0$~nm, which is reproduced here as Fig.~\ref{Fig6}(c).
There, the E1 band is located below the HH1 band, with a hybridization gap opening at $k_\parallel \neq 0$.
As $d_\mathrm{InAs}$ is decreased, the increased quantum confinement in the InAs layer raises the E1 level at $k_\parallel = 0$ and, at a critical InAs thickness $d_\mathrm{c}$ ($= 7.7$~nm), E1 coincides with HH1, where the system becomes gapless [Fig.~\ref{Fig6}(b)].
Upon further decreasing $d_\mathrm{InAs}$, the system enters the non-inverted regime, where E1 is located above HH1, separated by a normal gap at $k_\parallel = 0$ [Fig.~\ref{Fig6}(a)].
In Fig.~\ref{Fig6}(e), the energy levels at $k_\parallel = 0$ are plotted as a function of $d_\mathrm{InAs}$, which makes it clear that the topological phase transition occurs as a result of the level crossing between E1 and HH1 at $d_\mathrm{InAs} = d_\mathrm{c}$.
Figure \ref{Fig5}(e) also reveals an anticrossing between E1 and LH1 \footnote{Recently, this E1-LH1 crossover has been shown to impact the position of the Dirac point in the edge-state dispersion relative to the bulk energy gap~\cite{Skolasinski2018}}, which explains why the E1 band acquires an LH character near $k_\parallel = 0$ when $d_\mathrm{InAs}$ increases to $11$~nm [Fig.~\ref{Fig6}(d)].

When CQWs are pseudomorphically grown on AlSb substrates and the GaSb layer is tensile-strained, the $d_\mathrm{InAs}$ dependence of the band structure becomes significantly different [Figs.~\ref{Fig6}(f)-(i)].
For the thinnest $d_\mathrm{InAs}$ ($= 7.5$~nm), the system is in the non-inverted regime with E1 located above HH1 [Fig.~\ref{Fig6}(f)].
Although the dispersion looks similar to that of thin CQWs on GaSb substrates, a notable difference is that the band gap is now indirect; i.e., it forms between the bottom of E1 at $k_\parallel = 0$ and the maximum of the HH1 band at $k_\parallel = 0.4$~nm$^{-1}$ (where the latter has an LH character).
As $d_\mathrm{InAs}$ increases, this indirect band gap closes when the E1 band moves down, and its bottom at $k_\parallel = 0$ coincides with the maximum of the HH1 band at $k_\parallel \neq 0$ [Fig.~\ref{Fig6}(g)].
We denote this $d_\mathrm{InAs}$ ($= 8.3$~nm) at which the indirect gap closes as $d_\mathrm{c}^\prime$, and distinguish it from $d_\mathrm{c}$ at which a direct gap closes as a result of band touching.
For $d_\mathrm{InAs} > d_\mathrm{c}^\prime$, E1 continues to lower and the indirect gap becomes negative, making the system a semimetal as we discussed in the previous section for $d_\mathrm{InAs} = 9.0$~nm.
As $d_\mathrm{InAs}$ increases further, the E1 and HH1 bands touch at $k_\parallel = 0$ [Fig.~\ref{Fig6}(h)], and then their order becomes inverted for $d_\mathrm{InAs} > d_\mathrm{c}$ [Fig.~\ref{Fig6}(i)].
Even though there is no band touching for $d_\mathrm{InAs} > d_\mathrm{c}$, the system remains semimetallic because the LH-like band around $k_\parallel = 0.4$~nm$^{-1}$ stays higher up in energy.
The evolution of the energy levels at $k_\parallel = 0$ as a function of $d_\mathrm{InAs}$ [Fig.~\ref{Fig6}(j)] is similar to that for CQWs on GaSb substrate, except that E1 is strongly mixed with LH1 over a wide range of $d_\mathrm{InAs}$ including $d_\mathrm{InAs} = d_\mathrm{c}$.
The topological phase transition, which must be accompanied by a closure of a direct gap (i.e., band touching), is found to occur at $d_\mathrm{InAs} = d_\mathrm{c}$.
We see that at $d_\mathrm{InAs} = d_\mathrm{c}$ HH1 is crossed by a level having an LH character.
It is important to note that this topological transition at $d_\mathrm{InAs} = d_\mathrm{c}$ is pre-empted by a transition to a semimetal at $d_\mathrm{InAs} = d_\mathrm{c}^\prime$ and could therefore be masked in transport measurements.

The above line of argument based on the band calculation as a function of $d_\mathrm{InAs}$ and strain is consistent with the results of the transport experiments presented in the preceding subsections.
The resistivity peak observed at the CNP of the unstrained-GaSb CQWs in the inverted regime indicates the opening of a hybridization gap.
The absence of such a resistivity peak at the CNP in the deeply inverted CQW with tensile-strained GaSb suggests a semimetallic band structure.
Concerning the critical InAs thickness at which the band inversion takes place, our experimental results indicate $d_\mathrm{c} < 9.1$~nm and $10 \le d_\mathrm{c} < 10.9$~nm for the unstrained- and tensile-strained-GaSb CQWs, respectively (Fig.~\ref{Fig4}).
Although these estimates for $d_\mathrm{c}$ are greater than the calculated ones (7.7 and 9.7~nm, respectively), the trend that tensile strain shifts the critical thickness to larger $d_\mathrm{InAs}$ is consistent.
We note that our calculation underestimates $d_\mathrm{c}$ because it neglects the Hartree potential arising from the electron transfer from GaSb to InAs~\cite{Jiang2017, Andlauer2009}.
We emphasize, however, that the above discussion remains valid for understanding the impact of tensile strain and $d_\mathrm{InAs}$ at the qualitative level.

\subsection{$d_\mathrm{GaSb}$-$d_\mathrm{InAs}$ phase diagram}

\begin{figure}[ptb]
\includegraphics{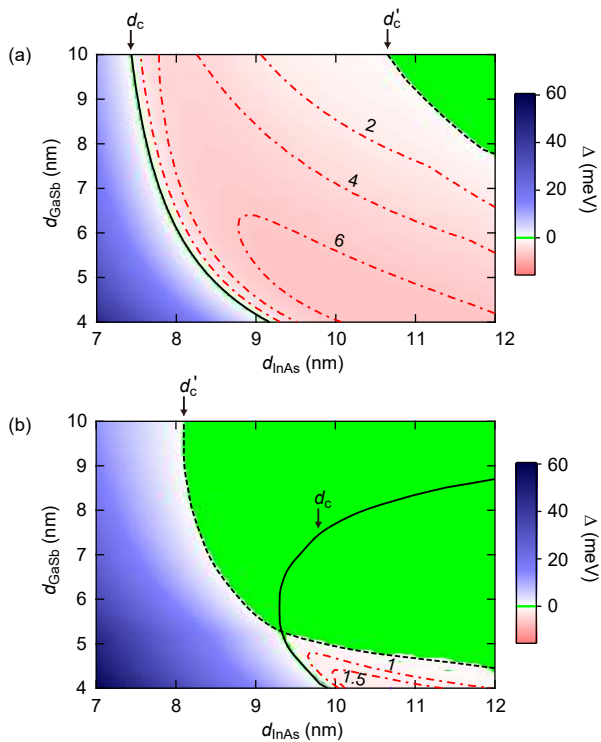}
\caption{(color online) Calculation of the band gap $|\Delta|$ as a function of $d_\mathrm{InAs}$ and $d_\mathrm{GaSb}$ for CQWs grown on (a) GaSb and (b) AlSb substrates.
The color maps show $\Delta$, with its positive and negative signs assigned to indicate a normal semiconducting gap in the non-inverted regime and hybridization gap in the inverted regime, respectively.
The green area indicating a gap closure over a wide parameter space represents a semimetallic phase.
The solid and dashed lines indicate the critical thicknesses $d_\mathrm{c}$ and $d_\mathrm{c}^\prime$, respectively.
The former separates the topological and nontopological regimes while the latter delineates the boundary of the semimetallic region.
The dash-dotted lines represent the contour line of the gap in the topological regime, with the labels showing the gap size $|\Delta|$ in meV.}
\label{Fig7}
\end{figure}

Our discussion so far has been confined to CQWs with a fixed $d_\mathrm{GaSb}$ ($= 7.3$~nm).
In the following, we examine the impact of $d_\mathrm{GaSb}$, another key parameter that dictates the band inversion and the size of the hybridization gap in the inverted regime~\cite{Liu2008, Skolasinski2018}.
We calculated the energy gap {$|\Delta|$} as a function of $d_\mathrm{InAs}$ and $d_\mathrm{GaSb}$.
Figures \ref{Fig7}(a) and \ref{Fig7}(b) show the results for CQWs pseudomorphically grown on GaSb and AlSb substrates, respectively.
These color maps show $\Delta$, to which we assign positive and negative signs in the non-inverted ($d_\mathrm{InAs} < d_\mathrm{c}$) and inverted ($d_\mathrm{InAs} > d_\mathrm{c}$) regimes, respectively, to distinguish the two gapped phases with $|\Delta| > 0$.
The green color represents regions with $\Delta = 0$, i.e., where the system is semimetallic.

The overall feature of the topological-nontopological phase diagram in Fig.~\ref{Fig7}(a) is similar to those in Refs.~\cite{Liu2008, Skolasinski2018}.
The phase boundary is defined by the line indicating $d_\mathrm{c}$ at each $d_\mathrm{GaSb}$, where the direct gap closes as a result of E1-HH1 band touching and $\Delta$ changes sign \footnote{The exact position of the phase boundary depends on the details of the calculation.
A tensile strain in the InAs layer shifts the phase boundary to smaller $d_\mathrm{InAs}$, whereas the Hartree potential due to interlayer charge transfer tends to counteract it~\cite{Jiang2017}.
The former (latter) is neglected in Ref.~\cite{Liu2008} (our calculation), which implies that the resultant $d_\mathrm{c}$ is overestimated (underestimated).}.
In Fig.~\ref{Fig7}, we also show in a contour plot the size of the gap $|\Delta|$ in the topological regime.
For a fixed $d_\mathrm{GaSb}$, $|\Delta|$ takes its maximum at $d_\mathrm{InAs}$ slightly larger than $d_\mathrm{c}$ and then decreases slowly upon further increasing $d_\mathrm{InAs}$ as a result of the reduced wave function overlap between E1 and HH1.
For a similar reason, the largest gap ($\sim 6$~meV) is attained at small $d_\mathrm{GaSb} < 6.5$~nm.
When both $d_\mathrm{InAs}$ and $d_\mathrm{GaSb}$ are large, the system enters a semimetallic phase, where the strong LH character of the E1 band leads to an indirect gap closure similar to that shown in Fig.~\ref{Fig6}(g) \footnote{This semimetallic phase has been identified in the $d_\mathrm{InAs}$-$d_\mathrm{GaSb}$ phase diagram in Ref.~\cite{Skolasinski2018}}.

In contrast, the phase diagram for CQWs on AlSb substrate is significantly different [Fig.~\ref{Fig7}(b)].
Now the semimetallic phase prevails over a wide region, taking up a large portion of the $d_\mathrm{InAs}$-$d_\mathrm{GaSb}$ space that corresponded to the topological phase for CQWs on GaSb substrate.
As already explained, this happens because the tensile strain in the GaSb layer raises the energies of bands having an LH character, shifting the thickness $d_{c}^{\prime}$ at which the indirect gap closes to smaller $d_\mathrm{InAs}$.
The topology of each band, on the other hand, is expected to change at $d_\mathrm{InAs} = {d}_\mathrm{c}$, where the direct gap closes as a result of the level crossing between HH1 and E1/LH1.
As shown by the black solid line in Fig.~\ref{Fig7}(b), $d_\mathrm{c}$ first decreases with increasing $d_\mathrm{GaSb}$, but then starts to increase for $d_\mathrm{GaSb} > 6.5$~nm \footnote{The latter behavior stems from the strongly mixed character of the E1/LH1 band at $k_\parallel = 0$.
That is, without E1/LH1 mixing, the LH1 level rises more quickly than the HH1 level with increasing $d_\mathrm{GaSb}$.
To keep the E1/LH1 level aligned with the HH1 level, the E1 contribution to the energy must be reduced by increasing $d_\mathrm{InAs}$.}.
Consequently, for $d_\mathrm{GaSb} > 5.3$~nm, the semiconductor-to-topological-insulator transition expected at $d_\mathrm{InAs} = d_\mathrm{c}$ is pre-empted by a semiconductor-to-semimetal transition at $d_\mathrm{InAs} = d_\mathrm{c}^\prime$ ($< d_\mathrm{c}$) as we have seen in Figs.~\ref{Fig6}(f)-(i).
Therefore, despite the inverted band order and the direct-gap opening at $d_\mathrm{InAs} > d_\mathrm{c}$, the system is a semimetal.
Accordingly, a topological insulating phase exists only for $d_\mathrm{GaSb} < 5.3$~nm, where a strong quantum confinement of holes pushes down the LH1 band well below the HH1 band at $k_\parallel = 0$, making the band structure similar to that on the GaSb substrate.
Note, however, that only a tiny gap, much smaller than in Fig.~\ref{Fig7}(a), can be attained in this case.

\section{Discussion}

The phase diagram in Fig.~\ref{Fig7}(b) suggests that, for the tensile-strained-GaSb CQW with $d_\mathrm{GaSb} = 7.3$~nm used in our experiment, a topological insulating phase does not exist, having been taken over by a semimetallic phase.
As already pointed out, our calculations neglect the Hartree potential due to interlayer charge transfer, which leads to the overestimation of band inversion.
Accordingly, it is possible that in reality the semimetallic phase occupies a smaller portion of the $d_\mathrm{InAs}$-$d_\mathrm{GaSb}$ plane.
As shown in Fig.~\ref{Fig3}(c), the tensile-strained-GaSb CQW with $d_\mathrm{InAs} = 10.9$~nm clearly shows band inversion with $n_\mathrm{cross} = 1.0 \times 10^{15}$~m$^{-2}$, yet exhibits a resistivity peak reaching $63$~k$\Omega/\Box$ near the CNP.
Such a resistivity peak near the CNP is not expected from the simple two-carrier model, suggesting an opening of an energy gap.
The semiconductor-semimetal direct transition suggested in Fig.~\ref{Fig7}(b) would thus be relevant in CQWs with a wider GaSb well.
We add that our calculations neglect the in-plane band anisotropy due to the valence-band warping, which could be the dominant mechanism for the semimetallic behavior when the E-LH mixing is weak, i.e., in CQWs with an unstrained or narrow GaSb well.

Finally, we mention the relation between our results and those of Tiemann \textit{et al.}~\cite{Tiemann2017}.
The focus of Ref.~\cite{Tiemann2017} is on the effects of externally applied uniaxial strain, whereas we focus on the biaxial epitaxial strain due to lattice mismatch.
Nevertheless, the tight-binding calculations in Ref.~\cite{Tiemann2017} have shown that, even without strain, a deeply inverted CQW with 15-nm-thick InAs and 8-nm-thick GaSb is semimetallic, whereas an energy gap opens for a CQW with a narrower 12-nm-thick InAs well.
We note that their calculations assume InAs/GaSb CQWs pseudomorphically grown on GaSb so that the GaSb is unstrained in the absence of externally applied uniaxial strain.
In contrast, we consider a case where the GaSb well is under biaxial tensile strain, a situation that can occur when CQWs are grown on a thick Al(Ga)Sb buffer layer.
On the experimental side, Ref.~\cite{Tiemann2017} used a piezo device to externally apply uniaxial strain, the magnitude of which was 0.03\% at maximum.
To study the effects of epitaxial biaxial strain, we exploited different substrates and buffer layer structures, where the magnitude of the strain is much larger, i.e., up to 0.50\% (1.13\%) for GaSb (InAs).

\section{Summary}

We studied the influence of the epitaxial strain on the electronic properties of InAs/GaSb CQWs by magnetotransport measurements and eight-band $\mathbf{k\cdot p}$ calculations.
We have shown by both experiment and calculation that the tensile strain in the GaSb layer shifts the topological-nontopological phase boundary to a wider InAs well width.
In addition, our study reveals an adverse effect of tensile strain, namely the closing of the bulk gap resulting from the enhanced mixing of light-hole states into the heavy-hole band, which could explain why this system often behaves as a semimetal.
Our results thus give an insight into the heterostructure design for a robust QSHI state in InAs/GaSb CQWs, corroborating the importance of strain engineering as recently demonstrated for InAs/In$_{x}$Ga$_{1-x}$Sb CQWs with compressive strain~\cite{Akiho2016, Du2017, Li2017}.

\section*{Acknowledgement}
The authors thank H.~Murofushi for device fabrication.
This work was supported by the JSPS KAKENHI.
(No.~JP15H05854 and No.~JP26287068).

\appendix

\section{Strain evaluation method}

\begin{figure}[ptb]
\includegraphics{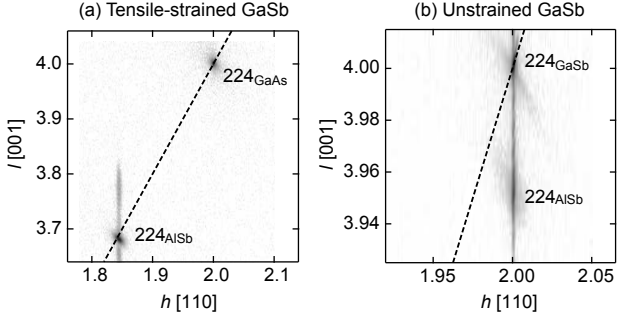}
\caption{Reciprocal space mapping of (a) tensile-strained- and (b) unstrained-GaSb CQWs.
The dashed line represents the expected peak position for the fully relaxed AlSb.}
\label{Fig8}
\end{figure}

We evaluated the magnitudes of strain in the CQWs from the two-dimensional reciprocal space mapping (RSM) of the high-resolution X-ray diffraction (HRXRD) at room temperature using GaAs(224) and  $(2 \overline{2} 4)$.
A clear 224$_\mathrm{AlSb}$ peak originating from the AlSb buffer was observed for both tensile-strained- and unstrained-GaSb samples.
As shown in Fig.~\ref{Fig8}(a), in the tensile-strained-GaSb sample, the reciprocal lattice point of the AlSb buffer layer is located on the dashed line connecting the origin and the 224$_\mathrm{GaAs}$ peak of the GaAs substrate, indicating that the buffer layer is almost fully relaxed.
The position of the 224$_\mathrm{AlSb}$ peak provides a relaxation ratio of 98\%, which is defined as $(a_\mathrm{buffer} -a_\mathrm{GaAs})/(a_\mathrm{AlSb} - a_\mathrm{GaAs})$, where $a_\mathrm{buffer}$ is the measured lattice constant of the AlSb buffer layer.
In contrast, in the unstrained-GaSb sample, the reciprocal lattice point of the thin AlSb buffer layer is vertically aligned with that of the GaSb substrate [Fig.~\ref{Fig8}(b)], indicating that the AlSb buffer layer is pseudomorphically strained.

\section{Equivalent-circuit model}

\begin{figure}[ptb]
\includegraphics{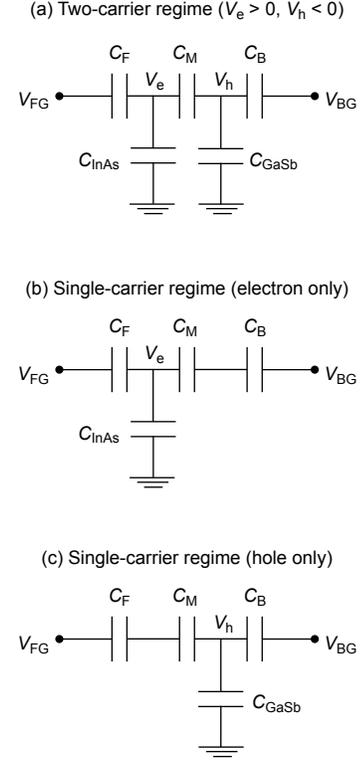}
\caption{Equivalent circuit for InAs/GaSb CQWs in (a) two-carrier regime, (b) single-carrier regime of electrons, and (c) single-carrier regime of holes.}
\label{Fig9}
\end{figure}

To describe the $V_\mathrm{FG}$ dependence of electron and hole densities, we use an equivalent-circuit model that takes into account the charge transfer and internal electric field in the QW layers.
As shown in Fig.~\ref{Fig9}(a), the equivalent circuit consists of three geometrical capacitances, which represent the couplings to the front and back gates ($C_\mathrm{F}$ and $C_\mathrm{B}$) and between the InAs and GaSb QW layers ($C_\mathrm{M}$), and two quantum capacitances of the QW layers ($C_\mathrm{InAs} = e^{2}m^{*}_\mathrm{e,InAs}/\pi \hbar^{2}$, where $m^{*}_\mathrm{e,InAs}$ is the electron effective mass of InAs, and $C_\mathrm{GaSb} = e^{2}m^{*}_\mathrm{h,GaSb}/\pi \hbar^{2}$, where $m^{*}_\mathrm{h,GaSb}$ is the hole effective mass of GaSb).
Voltage biases are applied through the front gate ($V_\mathrm{FG}$) and the back gate ($V_\mathrm{BG}$), while the QW layers are grounded.
The voltages $V_\mathrm{e}$ and $V_\mathrm{h}$ in Fig.~\ref{Fig9}(a) are related to the densities of electrons in InAs and holes in GaSb as $n_\mathrm{e} = C_\mathrm{InAs} V_\mathrm{e}/|e|$ and $n_\mathrm{h} = -C_\mathrm{GaSb} V_\mathrm{h}/|e|$, respectively.
Accordingly, electrons and holes exist when $V_\mathrm{e} > 0$ and $V_\mathrm{h} < 0$, respectively.

\begin{table*}
\caption {Parameters used in the equivalent-circuit model.
N.I. means that the sample is in the non-inverted regime.}
\begin{ruledtabular}
\begin{tabular}{ccccccccc}
Sample  &  $d_\mathrm{InAs}$ (nm)  &  $d_\mathrm{GaSb}$ (nm) &  $C_\mathrm{F}$ (nF/mm$^2$)  &  ${C}_{\mathrm{B}}$ (nF/mm$^2$) &  $m_\mathrm{e,InAs}$ ($m_0$) &  $m_\mathrm{h,GaSb}$ ($m_0$)  &  $n_\mathrm{cross}$ ($1/\mathrm{m}^{2}$) & $V_\mathrm{CNP}$ (V)\\
\hline
Tensile-strained & $11.8$ & $7.3$ & $0.84$ & $0.12$ & $0.09$ & $0.1$ & $2.1 \times 10^{15}$ & $-0.27$ \\
Tensile-strained & $10.9$ & $7.3$ & $0.75$ & $0.12$ & $0.09$ & $0.1$ & $1.0 \times 10^{15}$ & $-0.25$ \\
Tensile-strained & $10.0$ & $7.3$ & $0.75$ & $0.12$ & $0.09$ & $0.1$ & N.I. & $-0.11$ \\
Tensile-strained & $9.1$ & $7.3$ & $0.75$ & $0.12$ & $0.09$ & $0.1$ & N.I. & $-0.08$ \\
Unstrained & $10.6$ & $7.3$ & $0.82$ & $1.3$ & $0.06$ & $0.1$ & $2.4 \times 10^{15}$ & $-0.37$ \\
Unstrained & $9.1$ & $7.3$ & $0.88$ & $1.3$ & $0.09$ & $0.1$ & $0.6 \times 10^{15}$ & $-0.30$ \\
\end{tabular}
\end{ruledtabular}
\end{table*}

First, we consider the two-carrier (TC) regime, i.e., where $V_\mathrm{e} > 0$ and $V_\mathrm{h} < 0$.
The change in densities $\Delta n_\mathrm{e,TC}$ and $\Delta n_\mathrm{h,TC}$ in response to the change in gate voltages $\Delta V_\mathrm{FG}$ and $\Delta V_\mathrm{BG}$ can be written as
\[
\left[ \begin{array}{c}
\Delta n_\mathrm{e,TC} \\
\Delta n_\mathrm{h,TC} \end{array}
\right] = \frac{1}{|e|\mathrm{det}A}\left[ \begin{array}{cc}
C_\mathrm{InAs} & C_\mathrm{InAs} \\
-C_\mathrm{GaSb} & -C_\mathrm{GaSb} \end{array}
\right] A \left[ \begin{array}{c}
C_\mathrm{F} \Delta V_\mathrm{FG} \\
C_\mathrm{B} \Delta V_\mathrm{BG} \end{array}
\right],
\]
where
\[
A = \left[ \begin{array}{cc}
C_\mathrm{M} + C_\mathrm{B} + C_\mathrm{GaSb} & C_\mathrm{M} \\
C_\mathrm{M} & C_\mathrm{M} + C_\mathrm{F} + C_\mathrm{InAs} \end{array}
\right].
\]
These equations are modified to be suitable for fitting experimental data [Fig.~\ref{Fig2}(c)];
\[
n_\mathrm{e,TC}(V_\mathrm{FG}) = \frac{1}{|e|}\frac{C_\mathrm{InAs} (C_\mathrm{M} + C_\mathrm{B} + C_\mathrm{GaSb})}{\mathrm{det}A} C_\mathrm{F} (V_\mathrm{FG} -V_\mathrm{CNP}) + n_\mathrm{cross},
\]
\[
n_\mathrm{h,TC}(V_\mathrm{FG}) = -\frac{1}{|e|}\frac{C_\mathrm{GaSb} C_\mathrm{M}}{\mathrm{det}A} C_\mathrm{F} (V_\mathrm{FG} - V_\mathrm{CNP}) + n_\mathrm{cross},
\]
\[
n_\mathrm{net,TC}(V_\mathrm{FG}) = \left|n_\mathrm{e,TC}(V_\mathrm{FG})-n_\mathrm{h,TC}(V_\mathrm{FG})\right|,
\]
in which two fitting parameters, $V_\mathrm{CNP}$ and $n_\mathrm{cross}$, are introduced in such a way that $n_\mathrm{e,TC} = n_\mathrm{h,TC} = n_\mathrm{cross}$ at $V_\mathrm{FG} = V_\mathrm{CNP}$ for a fixed $V_\mathrm{BG}$ ($= 0$~V).
The equations above are valid only for the situation with $V_\mathrm{e} > 0$ and $V_\mathrm{h} < 0$ (i.e., $n_\mathrm{e,TC} > 0$ and $n_\mathrm{h,TC} > 0$).
Once we have $V_\mathrm{e} \le 0$ or $V_\mathrm{h} \ge 0$ in the studied $V_\mathrm{FG}$ range, the equivalent circuit needs to be replaced with that for the single-carrier (SC) regime hosting only electrons [Fig.~\ref{Fig9}(b)] or holes [Fig.~\ref{Fig9}(c)].
For the electron SC regime, the electron density can be written as
\[
n_\mathrm{e,SC}(V_\mathrm{FG}) = \frac{1}{|e|}\frac{C_\mathrm{InAs}}{C_\mathrm{F} + C_\mathrm{InAs} + C^\prime_\mathrm{B}} C_\mathrm{F} (V_\mathrm{FG} -V_\mathrm{FG1}) + n_\mathrm{e,TC}(V_\mathrm{FG1}),
\]
where $C^\prime_\mathrm{B} = {\left(\frac{1}{C_\mathrm{M}} + \frac{1}{C_\mathrm{B}}\right)}^{-1}$ and $V_\mathrm{FG1}$ is the front-gate voltage at the boundary between the TC and electron SC regimes.
Similarly, in the hole SC regime, the hole density can be written as
\[
n_{\mathrm{h,SC}}(V_{\mathrm{FG}})=-\frac{1}{\left|e\right|}\frac{C_{\mathrm{GaSb}}}{C_{\mathrm{B}}+C_{\mathrm{GaSb}}+C^{\prime}_{\mathrm{F}}}C^{\prime}_{\mathrm{F}}\left(V_{\mathrm{FG}}-V_{\mathrm{FG2}}\right)+n_{\mathrm{h,TC}}(V_{\mathrm{FG2}}),
\]
where $C^\prime_\mathrm{F} = \left(\frac{1}{C_\mathrm{F}} + \frac{1}{C_\mathrm{M}}\right)^{-1}$ and $V_\mathrm{FG2}$ is the front-gate voltage at the boundary between the TC and hole SC regimes.
By definition, the net carrier density in the electron (hole) SC regimes equals $n_\mathrm{e,SC}({V}_\mathrm{FG})$ [$n_\mathrm{h,SC}(V_\mathrm{FG})$].

Next, we describe how to determine the parameters used in our analysis [Figs.~\ref{Fig3}(c) and \ref{Fig3}(d)].
Table II shows the values used in Figs.~\ref{Fig3}(c) and \ref{Fig3}(d).
For inverted samples, $m^{*}_\mathrm{e,InAs}$ and $m^{*}_\mathrm{h,GaSb}$ were chosen so that the slope of the calculated $n_\mathrm{e,TC}(V_\mathrm{FG})$ matches that of experimental $n_\mathrm{SdH}(V_\mathrm{FG})$ in the two-carrier regime.
$m^{*}_\mathrm{e,InAs}$ was obtained as $(0.06\mathrm{-}0.09)m_{0}$, where $m_{0}$ is the free-electron mass.
These values are heavier than that at the band edge of bulk InAs ($0.024 m_{0}$) and those reported for InAs/GaSb CQWs ($0.04 m_{0}$~\cite{Qu2015}, $0.032 m_{0}$~\cite{Mu2016}).
We note that the latter was measured in the deep electron regime.
The larger $m^{*}_\mathrm{e,InAs}$ we obtained in the coexistence regime is thought to be affected more strongly by the hybridization of electron and hole wave functions.
On the other hand, $m^{*}_\mathrm{h,GaSb}$ cannot be accurately determined because the calculated $n_\mathrm{e,TC}(V_\mathrm{FG})$ [or $n_\mathrm{h,TC}(V_\mathrm{FG})$] is insensitive to $m^{*}_\mathrm{h,GaSb}$, unless $m^{*}_\mathrm{h,GaSb}$ takes an unreasonably small value.
Therefore, we tentatively used $m^{*}_\mathrm{h,GaSb} = 0.1 m_{0}$ in our analysis, which is close to the values reported for InAs/GaSb CQWs ($0.09 m_{0}$~\cite{Qu2015}, $0.136 m_{0}$~\cite{Mu2016}).
$V_\mathrm{CNP}$ and $n_\mathrm{cross}$ were determined so that the overall behavior matches the experimental data.
Using the $m^{*}_\mathrm{e,InAs}$ and $m^{*}_\mathrm{h,GaSb}$ values above, the band overlap can be estimated as $E_{\mathrm{g}0} = \hbar^{2}\pi n_\mathrm{cross}(m^{*}_\mathrm{e,InAs} + m^{*}_\mathrm{h,GaSb})/m^{*}_\mathrm{e,InAs}m^{*}_\mathrm{h,GaSb}$~\cite{Knez2011}.
The estimated $E_{\mathrm{g}0}$ ranges from $3$ to $15$~meV for $n_\mathrm{cross} = (0.6 \mathrm{-} 2.4)\times 10^{15}$~m$^{-2}$, which is reasonable in comparison with the result of $\mathbf{k}\cdot\mathbf{p}$ calculation that gives the same $n_\mathrm{cross}$.
For non-inverted samples, neither $m^{*}_\mathrm{e,InAs}$ nor $m^{*}_\mathrm{h,GaSb}$ can be determined, and thus we tentatively used $m^{*}_\mathrm{e,InAs} = 0.09 m_0$ and $m^{*}_\mathrm{h,GaSb} = 0.1 m_0$.


%

\end{document}